\documentstyle[12pt,epsf]{article}  

\begin{document}
\flushbottom

\baselineskip0.212in

\title{\vspace{2.2cm} \Huge \bf SOC in a Class\\ of Sandpile Models\\ with Stochastic Dynamics}
\author{ S. L\"{U}BECK and K. D. USADEL \\
 \it Theoretische Physik and SFB 166 \\ 
\it Universit\"{a}t Duisburg, Lotharstra{\ss}e 1, 
47048 Duisburg, FRG}
\date{}
\maketitle

\begin{abstract}
{\small
\baselineskip0.085in
\noindent We have studied one-dimensional cellular automata  
with updating rules depending stochastically on the 
difference of the heights of neighbouring cells. The probability for toppling
depends on a parameter $\lambda$ which goes to one with increasing
slope, i.e. the dynamics can be varied continuously. 
We have investigated the scaling properties of the model using finite-size
scaling analysis. A robust power-law behavior is observed for the 
probability density of the size of avalanches in a certain range of $\lambda$
values.
The two exponents which determine the dependence of the
probability density on time and system size both depend continuously
on $\lambda$, i.e. the model exhibits nonuniversal behavior. 
We also measured the roughness of the surface
of the sandpile and here we obtained an universal behavior, i.e. a roughness exponent 
of about 1.75 for all values of $\lambda$.
For the temporal behavior of the mass a $f^{-\phi}$ spectrum is obtained
with an exponent $\phi$ close to $2$ again for all $\lambda$-values.
}
\end{abstract}

{\vspace{-17cm}
\small
Fractals, Vol.~1, No.~4 (1993) 1030-1036.
\vspace{16.8cm}
}

\vspace{1.4cm}
{\large \bf 1.~INTRODUCTION}\\

Bak, Tang and Wiesenfeld \cite{BTW} recently introduced the notion of 
self-organized criticality (SOC) as a framework to understand the 
dynamics of extended dissipative dynamical systems. The dynamics drive
the system into a state where relaxation processes occur over all
time and length scales. It was suggested \cite{BTW} that sandpiles
were a particularly clear example of a self-organized system.

Theoretical sandpile models have been investigated by several authors
\cite{BTW}-\cite{CHRISTENSEN}. 
They usually involve two kinds of steps. First a particle
is added at ramdomly choosen sites, then the particles are eventually 
distributed to neighboring sites according to dynamical rules.
Usually, these rules are deterministic: there is toppling of particles
or not depending on the local occupation of sites. However, real
sandpiles behave in a different fashion. In real sandpiles the 
particles have different shapes, various sizes and they can stick
together. As a matter of fact, large and very unstable slopes can 
develop with slides of particles in an apparently stochastic fashion:
a certain slope may or may not lead to a slide depending on a great
number of internal degrees of freedom like the above mentioned 
different shapes and sizes of the particles, the friction between
them, the local variations in packing etc. To incorporate some 
of these features into a theoretical
model we propose as a first step a model with dynamical rules
which are not deterministic: toppling of particles occurs with a 
probability function which is small if the slope to neighboring
sites is small and which increases to one with increasing slope.
In the present paper we report results of extensive simulations of
this model. We restrict ourselves to the one-dimensional case
which shows already an  interesting and nontrivial behavior.\\ \\

{\large \bf 2.~THE MODEL AND SIMULATIONS}\\

The model we consider is an one-dimensional \quad"nonlocal unlimited"\quad 
sandpile model \cite{KADA} with the following rules. We assume
integer heights $h_i$ at lattice sites ${i=1,2,3,...,L}$, where L is the
size of the system. The local slope $\sigma_i$  at the site $i$ is defined as
the height difference between two neighbouring cells, ${\sigma_i=h_i-h_{i+1}}$.
The dynamics obey the following rules: If the slope ${\sigma_i}$ 
exceeds a threshold value $\sigma_c$ an avalanche can be generated.
With a certain probability $p(\sigma_i)$ grains drop from site $i$ to
the neighbouring cell with
\begin{displaymath}
h_i \rightarrow h_i - n_i
\end{displaymath}
\begin{equation}
h_{i+j} \rightarrow h_{i+j} + 1  \qquad \mbox{for} \quad  j=1,2,...,n_i
\end{equation}
The number $n_i$ of grains which topple grows with increasing $\sigma_i$,
${n_i=\sigma_i-N}$ (unlimited model). 
Since there is toppling to $n_i$ right neighbours the model is
called nonlocal. Note that from  {Eq.(1)} it follows that all slopes in
the system are non-negative. 

The simulation starts by adding particles at the 
top until the stability condition
${\sigma_1<\sigma_c}$ is violated. Then, with probability
$p(\sigma_1)$, $n_1$ grains  flow to the right neighbour 
according to Eq.(1) and an avalanche develops. 
If no toppling takes place another grain is added to site 1.
If toppling takes place the next site will be visited and so on. An
avalanche stops at site $i$ if this site doesn't topple or
if its slope to site ${i+1}$ is too small. Thus an avalanche 
only starts at the top. The length of an avalanche is defined
as the number of sliding sites. At the right edge of the system
grains can leave the pile, i.e. $h_i$ is set equal to zero for ${i>L}$.

The most important new ingredients in our model is the probability
function $p(\sigma_i)$ which determines whether a certain site $i$
topples or not. This function should obey the following
conditions:
\begin{displaymath}
p(\sigma<\sigma_c) \equiv 0
\end{displaymath} 
\begin{equation}
p(\sigma^{'})>p(\sigma) \qquad \mbox{for} \quad \sigma^{'}>\sigma
\end{equation}
\begin{displaymath}
p(\sigma) \rightarrow 1 \qquad \mbox{for} \quad \sigma \rightarrow \infty
\end{displaymath}
The first condition states that there is only toppling if
a threshold value is reached, the second conditions means that
there is increasing probability for toppling if the slopes increase
and the third equation means that toppling always takes place
for very large slopes, i.e. the heights $h_i$ are finite for any
system of finite length. In our simulation we use
\begin{equation}
p(\sigma_i) = 1-e^{-\lambda (\sigma_i-1)}
\end{equation}
$\lambda$ is called the toppling parameter. Thus there are three
parameters in our model: $\sigma_c$ which determines the stability
condition, $N$ which determines the number of sliding grains in
one event and most importantly the toppling parameter $\lambda$
with which we can tune the dynamics continuously.
In our simulations we choose $\sigma_c=2$, $N=1$ and study in particular
the dependence of the model on $\lambda$.
Note that for ${\lambda \rightarrow \infty}$ one gets a trivial model: 
${\sigma_i=1}$ for all $i$ and any avalanche reaches the edge of the
pile. 

The most important quantity to study is the probability density for
the lengths of the avalanches
for which a finite-size scaling analysis is performed. As
usual we assume  for this probability density $P(t,L)$ a scaling form
\begin{equation}
P(t,L)=L^{-\beta} g(L^{-\nu}t)
\end{equation}
where t is the number of toppling sites.
For ${t \ll L}$ we expect a power-law behavior, 
${P(t,L) \sim t^{-\kappa}}$ independent of $L$.
Thus $\beta$,$\nu$ and $\kappa$ must obey the following scaling
relation 
\cite{KADA} 
\begin{equation}
\beta = \nu \kappa 
\end{equation}
Another important quantity is the roughness exponent $\zeta$ which a is
measure for the 
fluctuations about the average height profile and which reflects the critical
nature of the steady state. It is defined by \cite{KRUG}:
\begin{equation}
\xi = \sqrt {L^{-1} \sum_{i=1}^L {<(h_i-<h_i>)^{2}>\quad}} \sim L^{\zeta}
\end{equation}
Finally we measure the total mass of the pile
\begin{equation}
M(t) = \sum_{i=1}^L h_i(t)
\end{equation}
and subsequently calculated its Fourier spectrum ${|M(f)|^{2}}$ to
determine the exponent $\phi$. \\ \\

{\large \bf 3.~RESULTS}\\

\begin{figure}[t]
 \vspace{-1.0cm}
 \begin{center}
 \begin{minipage}[t]{15.6cm}
 \epsfysize=7.9cm
\hspace{2.5cm} \epsffile{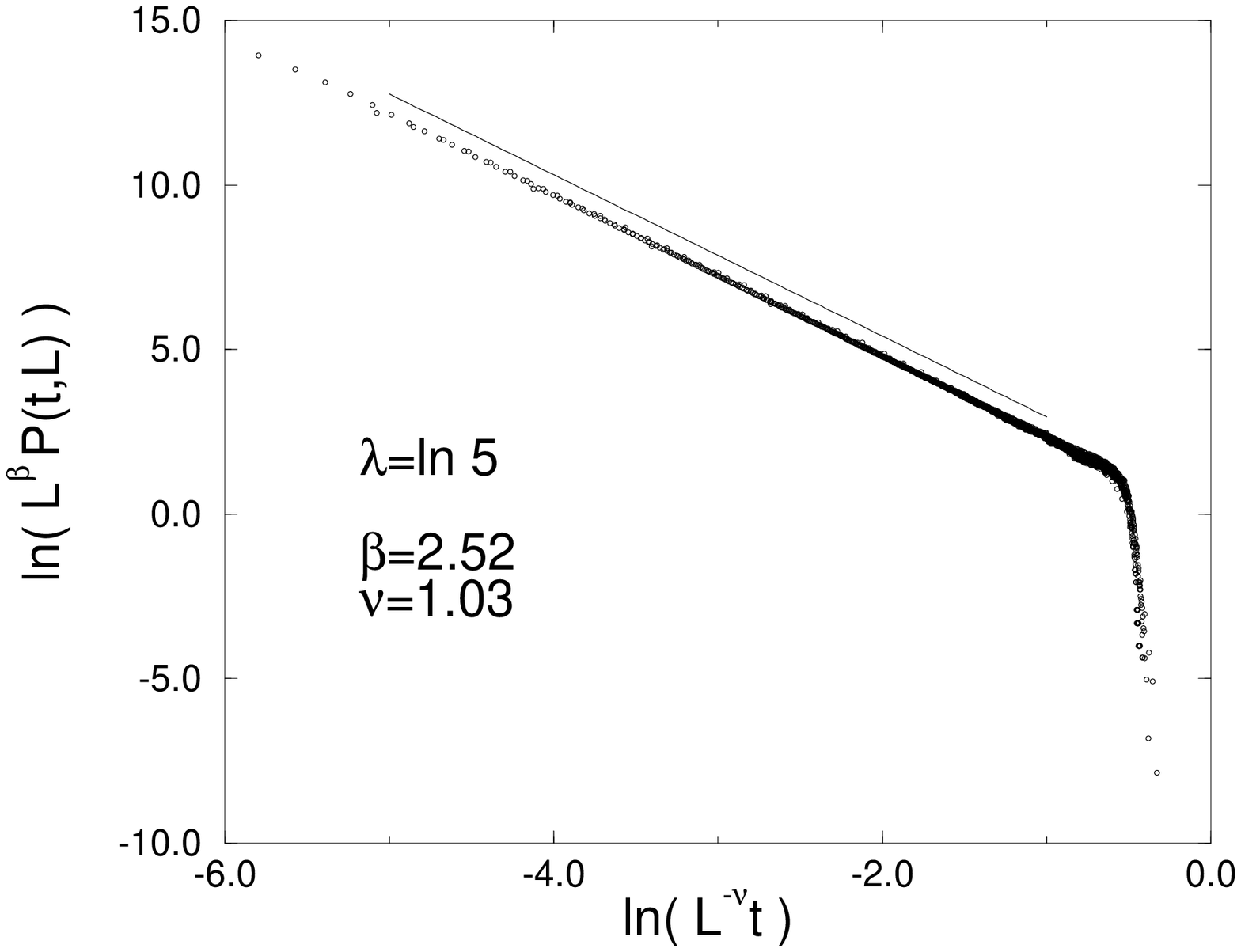} 
\vspace{-0.9cm}
 \caption{\small Finite-size scaling fit for $\lambda=ln\,5$ and five 
	  different lattice sizes
	  $L=50, 100, 200, 500, 1000$. The solid 
	  line corresponds to a power-law 
	  with an exponent $\kappa=2.453$.
 }
 \epsfysize=7.9cm
\hspace{2.5cm}  \epsffile{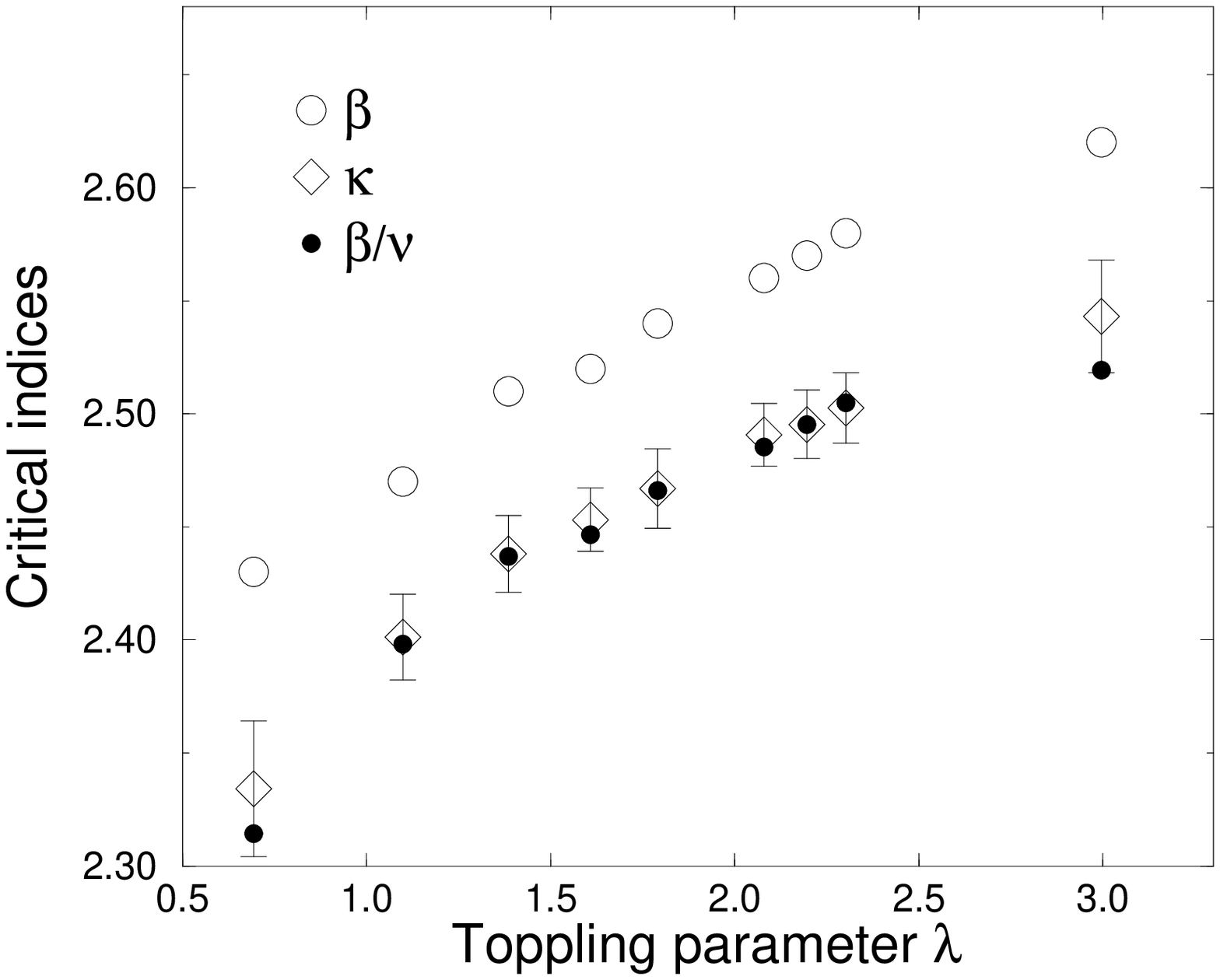} 
\vspace{-0.5cm}
 \caption{\small Scaling exponent $\beta$ and power-law exponent $\kappa$ as a function of
the toppling parameter $\lambda$. Solid circles correspond to ${\nu^{-1}\beta}$. 
Eq. (5) is fulfilled within numerical accuracy.	  
 }
 \end{minipage}
 \end{center}
\end{figure}

The simulations were performed for lattice sizes of
$L=50, 100, 200, 500, 1000$.
Before performing a measurement a large number of particles 
(arround $10^{8}$) was added starting 
from an empty lattice to reach the critical state.
Note that all measured values are independent of the initial conditions, i.e.
the critical state is an attractor of the dynamics.

First we measured the probability density of the length of the avalanches $P(t,L)$
for a lattice size $L=100$ and examined how $P(t,L)$ depends on the toppling 
parameter $\lambda$. A power-law behavior is observed in the 
range ${1<\lambda<3}$.
Outside this range we obtained on a log-log plot curves which deviate visible
from a
power-law behavior. We have not investigated this problem any further but
restricted ourselves  to the $\lambda$-values where power-law behavior is clearly
observed. Then the probability density  $P(t,L)$ was measured for various values
of $\lambda$ and lattice sizes. As can be seen from Fig.1 finite-size scaling  
works extremely well. For different values of $\lambda$ we get
${\nu \approx 1}$ but the scaling exponent $\beta$ changes strongly [Fig.2]. 
Thus we expect a varying 
exponent $\kappa$. $\kappa$ is calculated using regression analysis for 
lattice sizes $L=100,200,500,1000$. The averaged value of $\kappa$ is shown 
in Fig.2. Note that the change of the exponent can not be explained by 
statistical errors since they are far too small. The relation between the 
exponents, Eq.(5), is always fulfilled. 

\begin{figure}[t]
 \vspace{-1.0cm}
 \begin{center}
 \begin{minipage}[h]{15.6cm}
 \epsfysize=7.6cm
\hspace{2.9cm} \epsffile{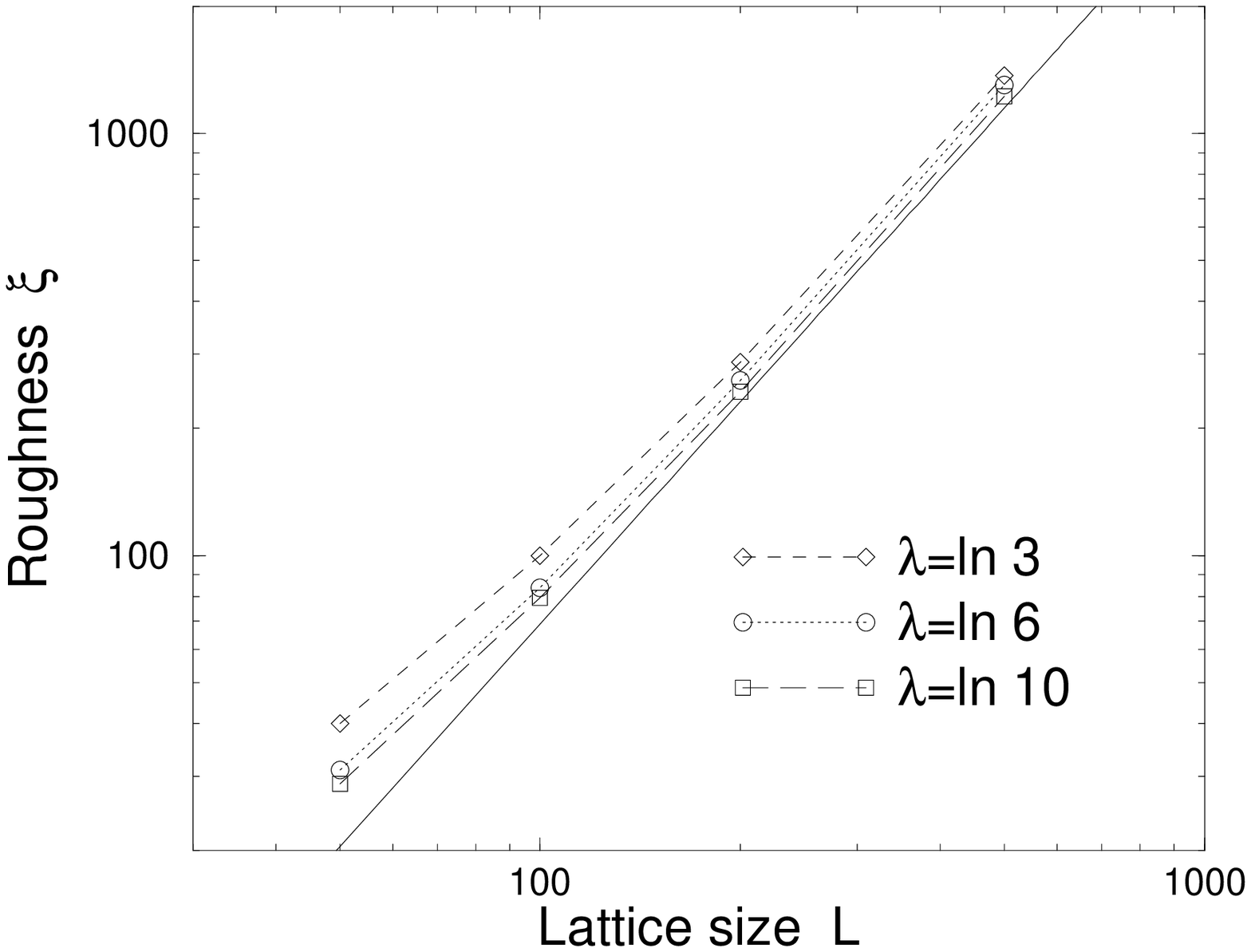} 
\vspace{-0.4cm}
 \caption{\small Roughness $\xi$ as a function of lattice size for different values of
the toppling parameter $\lambda$. The solid line corresponds to an power-law
${\xi\sim L^{\zeta}}$ with ${\zeta=1.75}$. 
 }
 \epsfysize=7.5cm
\hspace{2.9cm} \epsffile{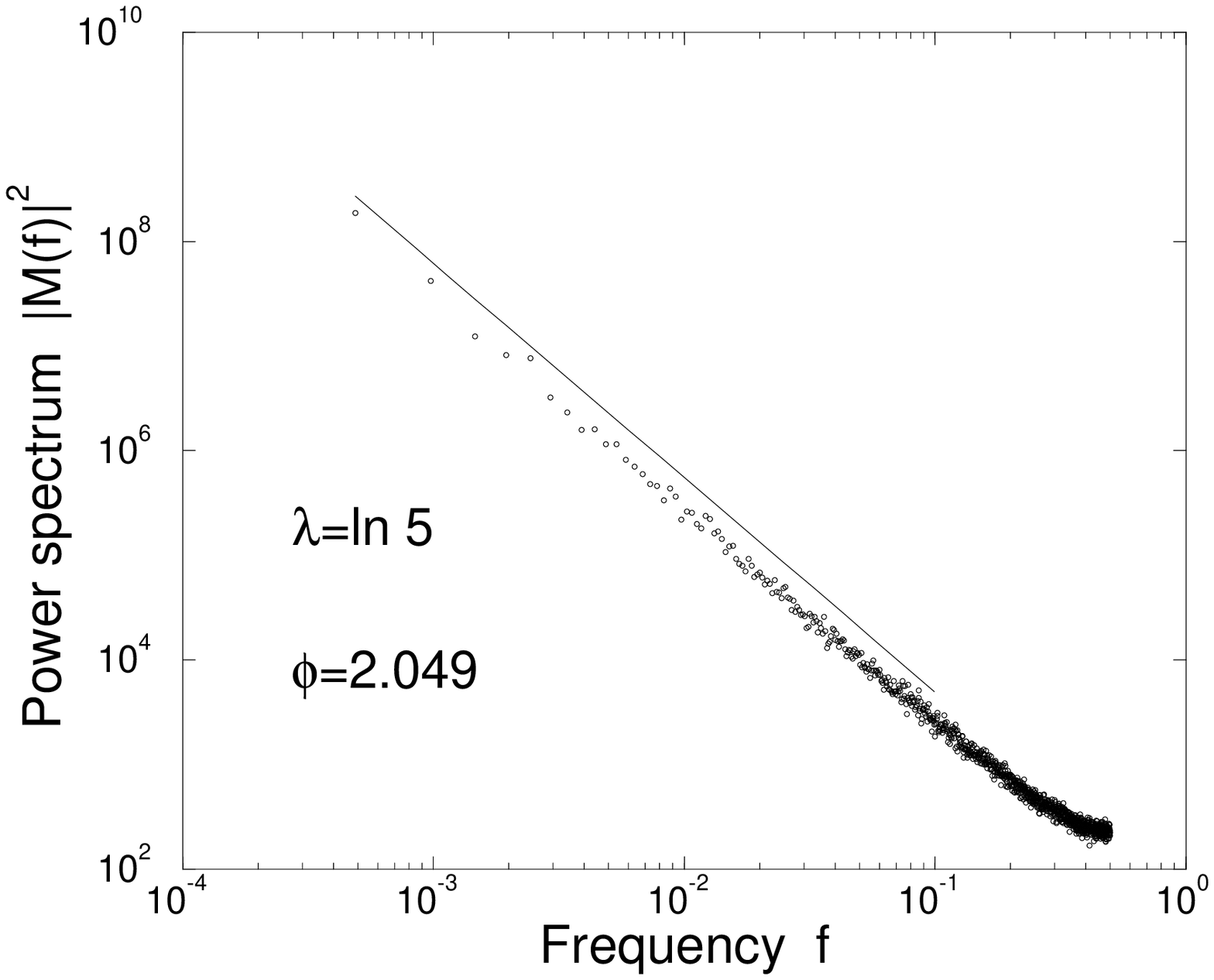} 
\vspace{-0.7cm}
 \caption{\small Power spectrum of the time evolution of the total mass of the pile for
${\lambda=ln\,5}$. The spectrum is an average over 40 measurements 
with 2048 events. Solid line corresponds to ${|M(f)|^{2}\sim f^{-2.049}}$.
 }
 \end{minipage}
 \end{center}
\end{figure}

To determine the roughness exponents $\zeta$ we limit our simulation to lattice 
sizes ${L \le 500}$. After reaching the critical state we add around
$10^{8}$ particles to calculate $\xi$ for 
different values of $\lambda$. Fig.3 shows the results. One can see that for 
${L \ge 100}$ the roughness $\xi$ scales with the system size as $L^{\zeta}$ and
$\zeta$ is independent of the toppling parameter $\lambda$. However,
one should note 
that for a very accurate determination of $\zeta$ one needs simulations with
system sizes larger than $500$ since the data points shown in Fig.3 are still
not on a straight line. 
But those measurements would demand more than $10^{9}$ 
events and have not yet been done. Increasing the lattice
size the roughness grows very fast, because ${\zeta \ge 1}$. The average height
profile scales as ${<h_i> \sim L^{\alpha_1}}$ with ${\alpha_1 \approx 1.8}$ 
for all $i$ and $\lambda$ \footnote{Notice that the average height profile is
not a linaer function of $i$, but obeys the scaling relation 
${<h_i>=L^{\alpha_1} f(L^{-\alpha_2}i)}$ with ${f= a_0+ a_1 e^{-a_2 i^{2}}}$ 
\cite{LUEBECK}.}. 
The exponents 
$\alpha_1$ and $\zeta$ should obey the relation ${\alpha_1 \ge \zeta}$ which 
is fulfilled in our model. It is very interesting to see whether the roughness
exponent $\zeta$ is equal to the scaling exponent of the averaged height 
$\alpha_1$.
Finally we have calculated the time dependence of the total mass of the pile (Eq.7).
We used a lattice size $L=50$ and measured the mass $M(t)$ for $2048$ events.
Fig.4 shows the average over 40 of the corresponding fourier-spectra. 
The power spectrum ${|M(f)|^{2}}$ scales as 
$f^{-\phi}$ with ${\phi \approx 2}$ for all $\lambda$. \\ \\

{\large \bf 4.~CONCLUSIONS}\\

\begin{center}
\begin{table}[t]
\vspace{-0.2cm}
\caption{\small The critical exponents for different values 
of the toppling parameter $\lambda$.}
\vspace{0.5cm}
\begin{center}
\begin{tabular}{lcccccc}
\hline
$\lambda$ & $\beta$ &  $\nu$ &  $\kappa$ &  $\zeta$ &  $\phi$  \\  
\hline
$ln\,2$  &  2.43  &  1.03  &  2.334  &  1.59  &  2.048 \\
$ln\,3$  &  2.47  &  1.03  &  2.401  &  1.70  &  2.061 \\
$ln\,4$  &  2.51  &  1.03  &  2.438  &  1.75  &  2.051 \\ 
$ln\,5$  &  2.52  &  1.03  &  2.453  &  1.80  &  2.049 \\
$ln\,6$  &  2.54  &  1.03  &  2.467  &  1.76  &  2.053 \\ 
$ln\,8$  &  2.56  &  1.03  &  2.491  &  1.75  &  2.064 \\ 
$ln\,9$  &  2.57  &  1.03  &  2.495  &  1.77  &  2.058 \\ 
$ln\,10$ &  2.58  &  1.04  &  2.503  &  1.76  &  2.112 \\ 
$ln\,20$ &  2.62  &  1.05  &  2.543  &  1.75  &  2.030 \\  
\hline
\end{tabular}
\end{center}
\end{table}
\end{center}
\vspace{-1cm}
In conclusion we have simulated an one-dimensional cellular automaton with
nonlocal, unlimited and stochastic dynamics which exhibits SOC behavior in 
a certain range of $\lambda$-values.
All trademarks of SOC are observed: finite-size scaling and a power-law behavior
of the avalanche probability as well as $f^{-\phi}$ behavior for the power spectrum.
However, our model displays nonuniversality:
the exponents $\beta$ and $\kappa$ change continuously with  $\lambda$. On the 
other hand the exponents which describe the fluctuations, $\phi$ and $\zeta$, are 
universal, i.e. independent of $\lambda$. Kert\'{e}sz et al \cite{KERTESZ} have 
shown that a great class of sandpile automata yield $f^{-2}$ behavior. 
Thus it is understandable why $\phi$ is constant. In contrast 
reasons why the roughness exponent is independent of the 
toppling parameter are not known to us. 
Generalization of this model to higher dimensions and an investigation of
the behavior for very small and very large values of $\lambda$ where 
power-law behavior apparently breaks down is under way.

\end{document}